\documentstyle[preprint,prb,aps]{revtex}  

\begin{document}                
\title{Collective Oscillations in Superconducting Thin Films 
in the Presence of Vortices}
\author{ Mauro M.~Doria$^a$, Fl\'avio M. R.~d'Almeida$^b$
 and O.~Buisson$^c$}
\address{
$^a$ Instituto de F\'{\i}sica, Universidade Federal Fluminense,
Campus da Praia Vermelha, Niter\'oi 24210-340 RJ, Brazil.\\
$^b$ Departamento de F\'{\i}sica, Pontif\'{\i}cia Universidade
Cat\'olica do Rio de Janeiro,  Rio de Janeiro 22452-970 RJ, Brazil.\\
$^c$ Centre de Recherches sur les Tr\`es Basses Temp\'eratures,
Laboratoire Associ\'e \`a l'Universit\'e\\Joseph Fourier,
C.N.R.S., BP 166, 38042 Grenoble-C\'edex 9, France.\\ }

\maketitle
\begin{abstract}
A plasma wave propagates inside an anisotropic superconducting film
sandwiched between two semi-infinite non-conducting bounding
dieletric media.
Along the c-axis, perpendicular to
the film surfaces, an external magnetic field is applied.
We show how vortices, known to cause dissipation and change the penetration
depth, affect the propagative mode.
We obtain the complex wave number  of  this mode and, using $YBCO$
at $4\;K$ as an example, determine a region  where
vortex contribution  is dominant  and dissipation is small.
\end{abstract}

\vskip 1.0truecm
\hskip 3.0truecm {\it PACS  \quad
74.20.De \quad 74.25.Nf \quad 74.76.Bz \quad 74.76.Db}

\newpage
\narrowtext
\section{Introduction}\label{section0}

In the plasma frequency, a collective oscillation of the electron gas
in the positive ionic background occurs, which is fundamental
to understand the electromagnetic properties of  conductors.
Below the plasma frequency  the conductor  reflects the incident
electromagnetic radiation and, above it becomes transparent thus allowing
a propagative mode.
For metals the plasma frequency is typically found in the ultraviolet
region ($10^{15}-10^{16}\; Hz$)\cite{PINES}.

A well-known feature of superconductors is the existence of a gap,
the energy required to break a Cooper pair in the ground state condensate. 
Typically the frequency associated to the gap for  conventional
superconductors is in the $10^{11}\;Hz$ range, whereas
 for the new high-Tc superconductors is one order magnitude higher
\cite{BEASLEY}.

The question whether the superconductor can support collective modes
 {\it without} inducing pair breaking effects
is an old one and has been discussed since the early days of the
theory of superconductivity \cite{MARTIN,ANDERSON}.
Apart from the so-called Carlson-Goldman mode,
which happens under special circumstances \cite{CARLSON,ARTEMENKO}, any other
 attempt to excite collective modes in isotropic {\it bulk} superconductors
 leads to the destruction of the superconducting state.
This follows the well-known argument\cite{ANDERSON} that the Coulomb
interaction shifts the frequency of such oscillations, the plasma frequency,
 to above the gap frequency.
However, it was recently shown that highly anisotropic 
 superconductors do display a plasma oscillation below the superconducting gap
\cite{TAMASAKU}.
This oscillation, specially to the layered structure, is due to the
 Josephson coupling between the superconducting planes.

Plasma modes in superconductors, isotropic or not,  has been recently 
revisited from another point of view.
Plasma modes below the gap are possible without destroying the superconducting
 state, as long as, they  propagate in the interface between the superconductor
 and a non-conducting bounding medium of very high dieletric constant.
This is the so-called superficial plasma mode \cite{RITCHIE,STERN},
which is made possible by the charges located at the interface of the
 superconductor and the dieletric medium, responsible for the creation of an
 electric field concentrated mainly outside the superconductor.

In a thin film, the coupling between the two superficial plasma modes
yields two possible branches, a symmetric and a anti-symmetric,
similarly to metals\cite{BOERSCH,FUKUI,SARID} and semiconductors\cite{UL}.
The film thickness must be smaller
than the  London penetration depth in order to produce this coupling.
Oscillations between the kinetic energy of the superelectrons
and the electrical field energy take place in these modes and for this reason
they are called {\it plasma modes}.
The lower frequency branch was predicted for superconductors
\cite{MOOIJ,MISHONOV,MIRHASHEN} some time ago
and  observed in   thin  granular aluminium films,
in the hundreds of $MHz$ range\cite{BUISSON},
and in thin  $YBa_2Cu_3O_{7-x}$ films\cite{DUNMORE}, in the
higher frequency range of hundreds of $GHz$.
The highest frequency branch is predicted to be within 
experimental observation range for the high Tc materials\cite{DPB}.
In case of highly anisotropic superconducting materials,
measurements of such upper and lower branches are expected to
give  information on the transverse and the longitudinal London penetration 
depths, respectively\cite{DPB}.
In conclusion, plasma modes in superconducting films can be an important tool
for the  probe of many intrinsic properties of superconductors.

Long ago Gittleman and Rosenblum\cite{GITTLEMAN}
have studied the effects of an applied
external current at the radio and microwave frequency range into
pinned vortices and obtained the surface impedance.
For an $AC$ applied magnetic field and in the weak pinning regime,
 Campbell\cite{CAMPBELL} showed that the effect of vortices can
 be described by a new $AC$ London penetration depth, whose square is 
 the original London penetration depth squared plus a new term describing
 the elastic interaction of vortices with the pinning centers.
A few years ago these models were generalized to convey the effects of
creep\cite{COFFEY} and to provide a more detailed description of the 
elastic properties of the vortex lattice near a surface\cite{BRANDT}.

In this paper we study the effects of a constant uniform magnetic field, 
applied perpendicularly to the thin film, into the thin film propagative 
mode. 
A sufficiently large magnetic field allows the thermodynamic stability 
of a vortex system, which influences the collective oscilations, 
affecting considerably the above modes.
The vortices are induced into an oscillatory dissipative motion around 
their pinning centers. 
This motion couples to the electromagnetic fields resulting either into an
underdamped or an overdamped regime.
This  paper is developed in the context of independent 
vortex and superelectron degrees of freedom.
We understand by  superelectron current, any supercurrent other than 
that one necessary to bring the thermodynamic equilibrium of vortices.
The vortex degree of freedom, described by its position in space, also
represents its intrinsic current.
In this framework arises the question whether the superelectron 
or the vortex contribution  dominates the propagative mode behavior.
Hereafter, by plasma mode we refer to the limit where superelectron
 contribution is the largest. 
So, pure plasma modes are found in the complete absence of an applied
magnetic field.
In this paper we discuss conditions that render the modes underdamped and 
vortex dominated.
This is the case of interest because the attenuated oscillations can be regarded
 as taking place between the vortex pinning energy  and the 
electrical field energy.

The present work is done  in  the  simplest possible
theoretical framework, essentially  a generalization
of the Gittleman-Rosenblum\cite{FIORY,KOSHELEV},
such that vortices and superelectrons are independently
coupled to  Maxwell's  theory. 
Here we are mainly interested in the low temperature regime and
therefore, ignore the contribution of normal electrons to the problem.
Thus,  wave damping is only due to the vortex dissipative motion.

We consider here an  anisotropic
superconductor with its uniaxial direction (c-axis) orthogonal to
the film surfaces: the  two London penetration depths,
transverse ($\lambda_{\perp}$) and longitudinal ($\lambda_{\parallel}$)
to the surfaces give an anisotropy such that
$\lambda_{\perp}/\lambda_{\parallel}>1 $ .
There are two dielectric constants, the non-conducting medium and  the
superconductor ones, $\tilde \varepsilon$ and $\varepsilon_s$, respectively.
Thus we are assigning to the superconductor a frequency independent dielectric
 constant.  We refer to the speed of light in the dielectric as
$v = c/\sqrt{\tilde \varepsilon}$.
The uniform  static applied magnetic field is $H_0$.
For each individual vortex, the viscous drag coefficient is $\eta_0$
and the elastic restoring force constant (Labusch parameter) is $\alpha_0$.
Their ratio, $\omega_0 \equiv \alpha_0/\eta_0$, is the so-called
depinning frequency, above which dissipation becomes dominant
in the vortex motion.
To have coupling between the two surfaces the film  thickness,  $d$,
must be  smaller than  $\lambda_{\parallel}$.

The choice of a nonconducting bounding medium of very high dieletric constant
 is crucial to lower the frequency range of the modes to below 
the gap frequency.
For this reason we take $SrTiO_3$  as the bounding media, whose dieletric 
constant is known to be high up to the $GHz$ frequency \cite{BUISSON}
 at low temperatures: $\tilde \varepsilon \approx 2.0\;10^{4}$.
Then the speed of light in the dielectric, $v=2.1\,10^6\; m s^{-1}$, is
 substantially smaller than $c$.
Our work is restricted to identical top and bottom dielectric
media, which does not imply lack of generality.
Similar  conclusions should also apply to the general asymmetric case.

This paper is organized as follows.
In the next section(\ref{section1})
we introduce the major equations  describing the film 
mode in the presence of vortices.
Its dispersion relation is  analytically derived under some justifiable
approximation.
In section \ref{section2} we apply our model
to the  high-Tc superconductor $YBa_2Cu_3O_{7-\delta}$,
 investigating a range of parameters such
that the lowest energy film mode is mostly associated to the vortex 
dynamics, but yet remains underdamped.
Finally, in section \ref{section3}, we summarize our major
results.

\section{ Propagating Modes in Superconducting Films with Perpendicular
Magnetic Field} \label{section1}

In this section we introduce the basic equations governing
wave propagation in
a superconducting film sandwiched between two identical
non-conducting dielectric media and subjected to an uniform static 
magnetic field  perpendicularly applied to the film surface.
An external electromagnetic wave of
angular frequency $\omega$ and vacuum wavenumber $k \equiv \omega/c$
is inserted in the dielectric bounded film.
We determine the dispersion relation of
 the lowest energy film mode, whose imaginary part reveals the
 attenuation behavior.
Phenomenological  theories, such as the present one, only describe the
superconductor in a energy range much lower than the pair breaking threshold.

 The electromagnetic dynamics of fields and superelectrons
is  described  by the Maxwell's equations,
\begin{eqnarray}
{\bf \nabla} \cdot \;\vec{D} &=& e\; (n_s-\bar{n}_s) \\
{\bf \nabla} \cdot\; \vec{H} &=& 0\\
{\bf \nabla} \times \;\vec{E} &=& -
\mu_0 {{\partial\vec{H}}\over{\partial t}}\\
{\bf \nabla} \times \; \vec{H} &=&  \vec{J} +
{{\partial\vec{D}}\over{\partial t}} \\
\end{eqnarray}
and consequently by the continuity equation,
\begin{eqnarray}
{\bf \nabla} \cdot \;\vec{J} + e\;{{\partial n_s}\over{\partial t}} = 0
\end{eqnarray}
where $n_s$ represents the space and time dependent charge density,
 $\bar{n}_s$ is its equilibrium value and $e$ stands for the electron charge.
The distinction between $n_s$ and $\bar{n}_s$ is necessary because, 
propagation through the system disrupts
the neutrality, as seen from Gauss' law ($n_s-\bar{n}_s=0$),
and the local charge density is no longer constant.

As previously  noticed the contribution of vortices and of superelectrons 
are independent in the present model.
The field $\vec{J}$, the superelectron current density involved in net 
macroscopic transport, and the field $\vec{u}$, the vortex displacement from
 its equilibrium position are independent in the present model.
Hence  the supercurrent density $\vec{J}$ corresponds to a macroscopic 
average  of the total superelectron motion, where the supercurrent 
necessary to establish each vortex averages to zero.
This approximation, valid for the present purposes, cannot give any
information on the supercurrent distribution surrounding each vortex line.

The simplest possible model that treats the response of
the vortices to the presence of a supercurrent  external to them is
the harmonic approximation of Gittleman and
Rosenblum\cite{GITTLEMAN},
\begin{eqnarray}
\eta_0 {{\partial \vec{u} }\over{\partial t}} + \alpha_0  \vec{u} =
\Phi_0 (\vec{J} \times \hat n)  \label{greq}
\end{eqnarray}
where $\hat n$ is a unit vector parallel to the flux lines mean direction.
>From its turn,  the displacements of vortices from their equilibrium
positions affect the propagating electromagnetic wave.
Fiory and Hebard\cite{FIORY} have considered this question and
found that besides the kinetic inductance due to the superelectrons,
 the moving vortices also contribute, producing an electric
field inside the superconductor.
\begin{eqnarray}
\vec{E} = \mu_0 \lambda^2 \cdot {{\partial\vec{J} }\over{\partial t}} -
\mu_0 H_0 (  {{\partial \vec{u} }\over{\partial t}}  \times \hat n)
\label{fheq}
\end{eqnarray}

The assumption  of  anisotropy yields a tensorial
London penetration depth.
\begin{eqnarray}
{\bf \lambda}  =
\pmatrix{ \lambda_{\perp}  & 0 & 0  \cr 0 & \lambda_{\parallel}& 0
\cr 0 & 0 & \lambda_{\parallel}  \cr } \quad
\lambda_{\perp} = \sqrt{{ {m_{\perp}}\over{\mu_0 \bar{n}_s e^2}}} \quad
\lambda_{\parallel} = \sqrt{{ {m_{\parallel}}\over{\mu_0 \bar{n}_s e^2}}}
\end{eqnarray}

We pick  a coordinate system where the two plane parallel surfaces
separating the superconducting film from the dielectric medium are
at $x=d/2$ and $x=-d/2$, such that $\hat n \equiv \hat x$ and propagation
is along the $z$ axis.
Vortex displacement is described by a vector field parallel to the surfaces,
$ \vec{u} = u_y \hat y \,+\, u_z \hat z$, with no orthogonal components 
to them ($u_x=0$).
According to symmetry arguments,
all fields for the present geometry can be expressed as
$F_i(x)\; \exp{[-\imath (q\,z-\omega\,t)]}$, where the wave number $q$ 
have yet to be determined.
 Because vortex motion is dissipative,
the wave's amplitude decays exponentially with distance,
and one obtains for the fields's expression 
$F_i(x)\; \exp{(q''\,z)}\;\exp{[-\imath (q'\,z-\omega\,t)]}$.
Then wave number is a complex number, $q=q'+\imath q''$.

Solving Maxwell's equations for the chosen geometry
gives two independent sets of field components,
the transverse electric (TE) and  the
transverse magnetic (TM) propagating modes.
In the former the non-zero electromagnetic field components are
$H_x$, $E_y$ and $H_z$, the non-vanishing supercurrent is $J_y$ and the
vortex displacement is along the
direction of wave propagation ($u_z$).
This is an extremely high frequency mode in the present theory,
and so not interesting because it lies above the gap.
For the latter, the non-zero electromagnetic field components are
$E_x$, $H_y$ and $E_z$, the non-vanishing supercurrent components are
$J_x$ and $J_z$ and the propagating wave displaces the vortices
perpendicularly to its direction of propagation ($u_y$).
This is a very interesting mode because it
supports low frequency propagating waves.
The major difference between TE and TM modes is that
the latter displays  superficial charge densities at
the film-dielectric interfaces and the former does not.
Such superficial charge densities stem from the
supercurrent component orthogonal to the film surface,
$J_x$, which is discontinuous at the interfaces, thus  rendering a
strong coupling between the superconducting film and the bounding media.

Introducing the time dependence $\exp{(\imath\;\omega\;t)}$
into Eq.(\ref{greq}) and Eq.(\ref{fheq}) results in a change of the 
penetration depth parallel to the surfaces
due to the vortex contribution\cite{BRANDT,CAMPBELL}.
\begin{eqnarray}
\imath \;\omega\;\mu_0 \lambda_{\perp}^2 \, J_x =  E_x , \qquad
\imath \;\omega\;\mu_0 \bar \lambda_{\parallel}^2 \, J_z =  E_z \\
\bar \lambda_{\parallel}^2 =  \lambda_{\parallel}^2 +
\big( {{B_0\;\Phi_0}\over{\mu_0\;\alpha_0}} \big)
{{1}\over{1+ \imath (\omega / \omega_0)}}  \label{lbar}
\end{eqnarray}
This equation shows that vortices and superelectrons contribute
additively to the parallel penetration depth.
Notice the depinning frequency $\omega_0$ establishes two distinct
 physical regions for the vortices response.
 For $ \omega \ll \omega_0 $  dissipation is weak and
$\bar \lambda_{\parallel}$ is essentially a real number.
For $ \omega \geq \omega_0 $ and a sufficiently large magnetic field,
dissipation  dominates the vortices response because
 $\bar \lambda_{\parallel}$ is  complex. 

The  superconductor's  dielectric constant is
tensorial, $\vec{D} =  \epsilon_{0}\varepsilon_{s} \vec{E} -\imath \vec{J}/\omega =
\epsilon_0 {\bf \varepsilon} \cdot \vec{E} $ and for the TM mode we have
that
\begin{eqnarray}
\varepsilon_{x}= \varepsilon_{s}- {1 \over {(k\lambda_{\perp})^2}} , \quad
\varepsilon_{z}=\varepsilon_{s}-{1\over {(k\bar \lambda_{\parallel})^2}},
\quad k \equiv {{\omega}\over{c}}
\end{eqnarray}

The TM field equations for the dielectric medium,
 ($x \ge d/2$ and $x\le -d/2$), are given bellow:
\begin{eqnarray}
E_x = \imath {{q} \over {\tilde \tau^2 }}
\;{{\partial E_z}\over{\partial x}}, \quad
H_y = \imath \,\epsilon_0\,{{\omega\, \tilde \varepsilon}\over{\tilde \tau^2}} \;
{{\partial E_z}\over{\partial x}}, \quad
{{\partial^2 E_z}\over{\partial x^2}} - {\tilde \tau}^2\, E_z = 0 ,
\quad {\tilde \tau}^2 = q^2 - k^2  \tilde \varepsilon \label{eqdie}
\end{eqnarray}
and the ones for the superconducting film ($ -d/2 \le x \le d/2$) follow. 
\begin{eqnarray}
E_x = \imath{{q \, \varepsilon_z  } \over
{\tau^2 \, \varepsilon_x }} \;{{\partial E_z}\over{\partial x}}, \quad
H_y = \imath \, \epsilon_0\,{{\omega\, \varepsilon_z}\over{\tau^2}} \;
{{\partial E_z}\over{\partial x}}, \quad
{{\partial^2 E_z}\over{\partial x^2}} - \tau^2\, E_z = 0
\quad \tau^2 = {{ \varepsilon_z}\over{\varepsilon_x}}q^2
- k^2  \varepsilon_{z}  \label{eqsuc}
\end{eqnarray}

The dispersion relations follow from the continuity of the
ratio  $H_y/E_z$ at a single interface, say $x=d/2$, once
assumed the longitudinal field $E_z$ has a definite symmetry.
It happens in this way because, the superconductor film is bounded by the same
dielectric medium in both sides.
Solving Eq.(\ref{eqdie}) one gets that
above  the film ($x \ge d/2 $),
\begin{eqnarray}
E_z = \tilde E_o \exp{(-\tilde \tau\;x)} \quad and \quad
{{\tilde H_y}\over{\tilde E_z}}|_{x=d/2} = -\imath {{\omega \epsilon_0
\tilde \varepsilon}\over{\tilde \tau}}
\label{ex}
\end{eqnarray}
>From Eq.(\ref{eqsuc}) we learn that for the superconducting film
 ($ -d/2 \le x \le d/2$) there are two
possible states, symmetrical and anti-symmetrical,
where the longitudinal field is expressed  by
$E_z = E_o \cosh{(\tau\; x)}$ and
$E_z = E_o \sinh{(\tau\; x)}$, respectively.
As discussed earlier, we shall only consider the symmetric branch,
the lowest mode in energy.
So the ratio of the tangential fields becomes
$H_y/ E_z|_{x=d/2} = \imath \omega \epsilon_0
\tilde \varepsilon_z \tanh{(\tau d/2)}/ \tau $.
Continuity of this ratio across the interface gives
 the following implicit relation.
\begin{eqnarray}
{{\tau\;\tilde\varepsilon}\over{\tilde\tau\;\varepsilon_{z}}}
= -\tanh{(\tau\;{d\over 2})} \label{tab1}
\end{eqnarray}
To find the dispersion relation we must solve Eq.(\ref{tab1}).
Here we use an approximate method to analytically solve it.
This approximation amounts to  replace the function
$(\tanh{z})/z$ in Eq.(\ref{tab1}), by another function,
$1/\sqrt{1+(2/3)z^2}$, which has an extremely close  behavior.
For $ z \ll 1$  both functions coincide up to the second order term
 in the Taylor series expansion: $1-(1/3)z^2+\dots$.
As all our results are derived in the range $z \ll 1$ thus, we replace 
 Eq.(\ref{tab1}) by the following approximate dispersion relation.
\begin{eqnarray}
{{\tilde \varepsilon}\over{\tilde \tau}} \approx - {{d \varepsilon_z}\over{2}}
{{1}\over{ \sqrt{1+{2\over3}\big( {{\tau d}\over{2}} \big)^2}}}
\label{drap1}
\end{eqnarray}
Squaring  the above expression, one obtains  a linear equation for $q^2$:
\begin{eqnarray}
q^2 = \big( {{\omega}\over{v}}\big)^2 {{
1 + \big({{2\omega \bar\lambda_{\parallel} }\over{d v}} \big)^2
\lbrack  \bar\lambda_{\parallel}^2+ {{d^2}\over{6}} \rbrack}\over{
1 - {2\over 3}\big({{\omega}\over{v}} \big)^4
\bar\lambda_{\parallel}^2 \lambda_{\perp}^2}}
\label{drap2}
\end{eqnarray}
The term proportional to $d^2/6$ in the numerator is irrelevant,
assuming the film much thinner than the penetration
depth ($\lambda_{\parallel} \gg d$).
We restrict the present study to frequencies much below  the 
assymptotic frequency ($({{\omega}\over{v}} \big)^4 \bar\lambda_{\parallel}^2
\lambda_{\perp}^2 \ll 1$), thus obtaining the following
dispersion relation: 
\begin{eqnarray}
q^2 = \big( {{\omega}\over{v}}\big)^2
(1 + \big({{2\omega \bar\lambda_{\parallel}^2} \over{d v }}\big)^2)
\label{drap3}
\end{eqnarray}
In  the absence of an applied uniform magnetic
field ($H_0=0$), consequently with no vortices,
there is no dissipation and $q''=0$.
In this case we retrieve the well-known dispersion relation of plasma modes
taking into account the retardation effect \cite{MIRHASHEN,BUISSON}.

Next we study two different behaviors of the dispersion relation
in the presence of vortices.

\vskip0.5truecm
\par \noindent $\underline{optical \quad mode}$ \quad
At low frequencies the  mode is, in leading order,
a plane wave travelling in the dielectric medium,
$q' \approx \omega /v$,
with no attenuation along the direction of propagation,
($q'' \approx 0$).
Perpendicularly to the film, the amplitude shows no attenuation,
because $\tilde \tau \approx 0$, according to Eq.(\ref{ex}).
We obtain, from  the  Taylor expansion of Eq.(\ref{drap2}),
the lowest order corrections in $\omega$ to the
above description of the optical regime.
\begin{eqnarray}
q' &=&  {{\omega}\over{v}} \lbrace
1+ {1\over 2} \lbrack  {{2\omega\big(
\lambda_{\parallel}^2 +
{{B_0\;\Phi_0}\over{\mu_0\;\alpha_0}}
\big)}\over{d v}} \rbrack^2 \rbrace  + \cdots \label{Kop} \\
q'' &=& - {{ 4 \omega^4 {{B_0\;\Phi_0}\over{\mu_0\;\alpha_0}} \big(
\lambda_{\parallel}^2 +
{{B_0\;\Phi_0}\over{\mu_0\;\alpha_0}} \big) }
\over{ v^3 d^2 \omega_0
}} + \cdots \label{Lop}
\end{eqnarray}
\vskip0.5truecm
\par \noindent $\underline{coupled \quad mode }$ \quad
For sufficiently large frequencies,
Eq.(\ref{drap3}) no more describes a linear response.
In this range the superconducting film and the dieletric media
are effectively coupled, which implies in a reduction of 
the mode propagation speed ($(\omega/q')/v \ll 1$).
This is the most interesting regime since  film and  dielectric 
produce a low energy  mode.

Far-way from the linear regime,
and provided that the asymptotic frequency is still out of range,
Eq.(\ref{drap3}) is approximately described  
by its second term, resulting into the dispersion relation
$ q \approx (2/ d) [\omega /(v \bar \lambda_{\parallel})]^2 $.
>From this, we obtain its wavevector and attenuation:
\begin{eqnarray}
q'(\omega) &=&  {{2\;\omega^2}\over{d\;v^2}} \; \lbrack
\lambda_{\parallel}^2 +  {{B_0\;\Phi_0}\over{\mu_0\;\alpha_0}}\;
{{1}\over{1+ \big(\omega/\omega_0\big)^2}} \rbrack  \label{Kcoup}\\
q''(\omega) &=& - {{B_0\;\Phi_0}\over{\mu_0\;\alpha_0}}
{{\omega^3/\big(\omega_0 v^2 d\big)}\over
{1+ \big(\omega/\omega_0\big)^2}}\label{cap}
\end{eqnarray}
In the frequency range where the above dispersion relation is a valid 
approximation, 
the ratio between the real and the imaginary
parts of  the London penetration depth, 
determines  whether the mode is overdamped or
underdamped:  
$q'/q'' = - Re(\bar \lambda_{\parallel}^2)/ Im(\bar \lambda_{\parallel}^2)$.

The cross-over  magnetic field,
\begin{eqnarray}
B_1 \equiv {{ \lambda_{\parallel}^2 \;\ \mu_0\;\alpha_0}\over{\Phi_0}}
\end{eqnarray}
splits  the regimes of  superelectron
($B_0 \ll B_1$), and vortex ($B_0 \gg B_1$) dominance.
 In these limits Eq.(\ref{lbar}) can be  replaced by
approximated expressions,
$\bar \lambda_{\parallel}^2 \approx  \lambda_{\parallel}^2$, and
$\bar \lambda_{\parallel}^2 \approx (B_0\;\Phi_0 /
\mu_0\;\alpha_0) / (1+\imath(\omega / \omega_0)) $, respectively.
Recall the assumption of the present model  that the superelectron 
contribution is never dissipative.
If in addition to an applied magnetic field much larger than
$B_1$,  we choose a frequency range  $\omega<\omega_0$, then $q''<q'$ and 
the mode is underdamped.
The dispersion relation follows a square root dependence, and becomes,
\begin{eqnarray}
\omega^2 \approx {{d v^2 \mu_0 \alpha_0} \over{2 B_0 \phi_0}} q' \label{cap2}
\end{eqnarray}
In this interesting limit the mode  energy shows many
oscillations between vortex and electric field energies before dissipation
dominates.
At higher frequency ($\omega>\omega_0$) this is no longer possible, 
since the mode becomes
overdamped due to the large dissipation of vortices above the depinning
frequency.
The frequency  $\omega_0$ coarsely defines a cross-over region
between the underdamped and the overdamped regimes.

In the next section, using experimental parameters measured on $YBCO$, 
we search for favorable conditions in frequency and magnetic field 
to observe underdamped coupled modes on a thin film.

\section{YBCO Thin Film}\label{section2}

In this section a
$YBa_2Cu_3O_{7-\delta}$  thin film  is taken,
as an example, to
determine a frequency and magnetic field window
where the  mode is coupled, underdamped
and vortex dominated.
The wave must be underdamped in order to travel over many wavelengths before its
amplitude is completely attenuated.
For this high-Tc superconductor the anisotropy
($\lambda_{\perp}/\lambda_{\parallel}=5$) and the zero-temperature
London penetration depth along the $CuO_2$ planes  are  well-known.
\cite{BLATTER,BRANDT2}.
At very low temperature several experiments
\cite{GOLOSOVSKY} have determined the viscosity and
the Labusch constant, all giving the same numbers,
which are summarized in table I.
Such parameters have a temperature dependence\cite{ANLAGE},
not taken into account here
because we only consider a fixed low temperature, namely, $4\;K$.
The magnetic field dependence of the Labusch constant, known to
exist for high-Tc materials\cite{HANAGURI} and low-Tc ones\cite{KOBER},
is not considered either.
For this discussion we choose  the film thickness  $d=10\; nm$.
\vskip 1.0truecm
\begin{center}
Table I~: Properties of the high-Tc material $YBa_2Cu_3O_{7-\delta}$ at
T = $4\,K $ \\
\vspace{0.5cm}
\begin{tabular}{||c|c|c|c|c||}
$\alpha_0$ ($N/m^2$) & $\eta_0$ ($N\,s/m^2$)
& $\omega_0={{\alpha_0}\over{\eta_0}}$ ($10^9\quad rad/s$) &
$\lambda_{\parallel}$ ($\mu m$)
& $B_1$ ($T$) \\
\hline \hline
 $3.0\;10^5$
& $1.2\; 10^{-6}$ & $250$ & $0.15$  & $4.1$
\end{tabular}
\end{center}
\vskip 1.0truecm


Fig.\ref{fig1}  provides
a pictorial intuitive view of the wave propagation inside the superconducting
film for the TM symmetric propagating mode.
Dimensions are out of proportion in order to enhance
some of the most relevant features.
Only the electric field lines
inside the superconducting film
are shown.
The  superficial charges are also shown and, represents
the sources of this propagating electric field.
The electric field lines
show a very important feature of this wave\cite{DPB},
namely,  the supercurrent component along the wave propagation direction, 
$J_z$,  is dominant over $J_x$.
A magnetic field perpendicularly applied to the film surfaces produces
vortices, pictorially represented at the top surface.
The oscillatory displacement suffered by vortices, because of the
driving Lorentz force caused by $J_z$ (Eq.(\ref{greq})),
is also shown in this figure.

As previously discussed, the adequate choice of frequency and magnetic field
windows is fundamental to observe the lower energy mode.
We can distinguish several different regions within the
$B_0$ vs. $\omega $ diagram.
Fig.\ref{fig2} shows such  regions for $YBCO$, 
according to the above parameters. 
 Two cross-over lines separate this diagram in three different regions:
the optical regime, the underdamped coupled regime and the
 overdamped coupled regime.

The lower line in Fig.\ref{fig2}, called $\omega_{cr}$, 
separates the optical region from the coupled regions.
This cross-over line  is defined through Eq.(\ref{Kop}),
using as conditon that the second term becomes a non negligible
 fraction $\chi_1$  of the first term and so can no longer be ignored,
\begin{eqnarray}
\omega_{cr} = \sqrt{\chi_1 \over 2} {{d v} \over { \lambda_{\parallel}^2 +
{{B_0\;\Phi_0}\over{\mu_0\;\alpha_0}}  }}  \label{omcr}
\end{eqnarray}
We have arbitrarily chosen ten percent ($\chi_1=0.1$)
as our criterion for the optical mode boundary.

The upper line in Fig.\ref{fig2}, called $\omega_{d\pm}$, 
is related to the dissipation and separates
the underdamped  to the overdamped regimes. 
The criterion for dissipation is  the ratio $q'/q''$,
which for the coupled regime, is approximately given by
the ratio between the real and the imaginary part of the squared
penetration depth $\bar \lambda_{\parallel}^2$ (Eq.(\ref{lbar})),
according to Eq.(\ref{cap}).
Thus our second cross-over line is defined by
$Im (\bar \lambda_{\parallel}^2) = \chi_2 Re (\bar \lambda_{\parallel}^2)$
where $\chi_2$ is an arbitrary factor.
This condition gives a second degree equation for $\omega/\omega_0$,
$\chi_2 \lambda_{\parallel}^2 (\omega/\omega_0)^2 -
(B_0\Phi_0/\alpha_0\mu_0)(\omega/\omega_0) +
\chi_2 ( \lambda_{\parallel}^2 + B_0\Phi_0/\alpha_0\mu_0 ) = 0$,
whose solutions,  $\omega_{d\pm}(B_0)$,
form the upper and lower branches of a single curve that encircles
the overdamped regime area.
\begin{eqnarray}
{{\omega_{d\pm}}\over{\omega_0}} = {{B_0}\over{2\chi_2 B_1}} \pm
\sqrt{\big({{B_0}\over{2\chi_2 B_1}}\big)^2 -{{B_0}\over{B_1}}-1}
\label{ompm}
\end{eqnarray}
Therefore, the dissipative region demands a minimum applied field
 $B_2$ to exist, defined by the vanishing of the above square root:
\begin{eqnarray}
B_2 = 2\chi_2(\chi_2+\sqrt{1+\chi_2^2})B_1
\label{b2}
\end{eqnarray}
Hence the  two curves $\omega_{+}$ and $\omega_{-}$
have a common start at $(B_2,\omega_2)$,
where $\omega_2= (\chi_2+\sqrt{1+\chi_2^2}) \omega_0$,
and  approach the asymptotic lines 
 $(\omega_0/\chi_2)(B_0 /B_1)$ and $\omega_0 \chi_2 $, respectively.
For the diagram in Fig.\ref{fig2}, we have taken $\chi_2=0.5$ thus, 
obtaining  that $B_2 \approx 6.64 \; T$ and
$\omega_2 \approx 4,05\;10^{11}\; rad/s$.
The asymptotic lines become $\omega_{d+}/\omega_0 \rightarrow 2 (B_0/B_1)$
and $\omega_{d-}/\omega_0 \rightarrow 0.5$.

As indicated by the  $B_0$ vs. $\omega$ Fig.\ref{fig2} diagram,
the modes are optical for frequencies below the $\omega_{cr}$ line where
they are weakly affected by the superconductor properties and the vortex 
dynamics.
In this region and for $ B_0 \ll B_1$
the superelectron dominates over the vortex response
and, effectively, there are plasma modes.
For $ B_0 \gg B_1$, the $\omega_{cr}$ line
decreases inversely proportional to $B_0$.
Above the $\omega_{d-}$ line, and, at large magnetic fields, $B > B_2$,
the  modes become overdamped.
Thus  the interesting region lies above the $\omega_{cr}$ line  
and below the $\omega_{d-}$ line,
where the modes are underdamped coupled and vortex dominated.
In this intermediate region, dissipation should be small enough ($q' > q''$)
to allow wave propagation over some wavelengths before attenuation sets in.

All Figures discussed below were obtained using 
Eq.(\ref{drap3}) expression. The complex
wave number $q$ is then easily derived as a function of $\omega$.

Fig.\ref{fig3} shows the dispersion relation $\omega$ vs. $q'$ for
$B_0 = 0 T$ and  $B_0 = 20 T$.
In case of zero magnetic field, 
The frequency window considered in this  figure is
below the zero magnetic field optical-coupled crossover
($\omega_{cr} \approx 2,10\; 10^{11}\; rad/s$).
Indeed, the $B_0 = 0 T$ mode shows  a quasi linear dependence.
However for a magnetic field $B_0 = 20 T$ , the presence of vortices 
changes dramatically the dispersion relation.
The frequency  $\omega_{cr}$ has droped substantially,
according to this figure.
Below the mode is optical, similarly to the zero magnetic field case,
and above  the mode
is slow in comparison to the zero field one.
This effect clearly comes from the vortex overwhelming contribution
at this large magnetic field value.
In this  frequency window, the mode is 
underdamped until the frequency $\omega_{d-}$ is reached.
Above it turns to be overdamped.
In order to better estimate the attenuation,
we have plotted the ratio $q'/q''$ for the same
frequency window (Fig.\ref{fig4}).
Notice that $q'/q''$, obtained from Eq.(\ref{drap3}) and shown here, 
gives directly the mode attenuation, whereas Eq.(\ref{cap})
just provides an approximate criterion, used to define the  dissipative
curve $\omega_{d\pm}$ of Fig.\ref{fig2}.
Fig.\ref{fig4} shows that for $\omega_{cr}<\omega<\omega_{d-}$
the mode  propagates over various wavelengths before
its amplitude goes to zero.
According to Fig.\ref{fig4} $q'/q''$
diverges for low frequencies within the optical regime.
This  behavior is explained  recalling that 
all losses are caused by vortices and disappear at zero frequency.

The reduced speed, defined as
the ratio between the phase velocity and the speed of light in the dielectric,
$(\omega/q')/v$ is plotted in Fig.\ref{fig5}.
In the optical regime this ratio is essentially equal to one.
This is quite verifiable at zero magnetic field but not at $20 T$, where
the modes are strongly slowered by the presence
of vortices.

\section{Conclusion}\label{section3}

In this paper  we have studied superficial coupled modes in
a superconducting film surrounded by two
identical dielectric media with an applied magnetic field
perpendicular to the surface.
The superconductor is anisotropic and its uniaxial
direction (c-axis) is perpendicular to the interfaces
with the dielectric medium.
The choice of non-conducting media of high dielectric constant
helps to lower the propagating wave frequency range much below
the gap frequency.
We consider a static magnetic field above the lowest critical field,
that allows for the existence of pinned and dissipative vortices.
In the present approach  superelectrons and vortices
contribute additively to the impedance.
Vortices and superelectrons interact with each other through
the Lorentz force and through an electric field, created by vortex
motion and superelectrons acceleration.
Here we have studied how the lowest energy branch,
the TM symmetric mode, is affected by vortices.
Under a justifiable approximation, we obtain an analytical expression
for this dispersion relation, which can describe simultaneously 
the three different possible behaviors for a propagating mode
in a superconducting film subjected to an exterior magnetic field,
namely, optical regime, underdamped coupled regime and overdamped 
coupled regime.

We find that in very high magnetic field,  vortices  
dominate over the superelectrons response.
The modes are well described by the vortex oscillations around their
pinning centers where, their energy oscillates
between the pinning energy and the electrical one.
We have studied the $B_0$ vs. $\omega$ diagram  
for a very thin superconducting
film, made of the high-Tc material $YBCO$. 
We find three different regions: 
optical, underdamped coupled and overdamped modes.
Nested between the optical and the overdamped regions, and above a certain
critical magnetic field cross-over, is
the region of interest.
There exist, in this frequency and magnetic field window,
underdamped  propagative modes,
whose behavior is determined by the vortex response, 
and not by the superelectrons.

This work was done under a CNRS(France)-CNPq(Brasil) collaboration
program.

\newpage

\newpage


\newpage

\baselineskip = 2\baselineskip  
\begin{figure}
\caption{ A pictorial view of wave propagation in a superconducting film
surrounded by identical non-conducting media in both sides.
Scales are out of proportion in order to enhance some of the features.
The instantaneous electric field is shown here only inside the film.
The superficial charge densities and the motion of the vortex lines are also
sketched here.}
\label{fig1}
\end{figure}
\begin{figure}
\caption{
The diagram $B$ vs. $\omega$ for
a very thin $YBCO$ superconducting film, $d=10nm$-thick, surrounded
by the dielectric material $SrTiO_3$ shows three regions:
optical, underdamped coupled and overdamped modes.
The dashed line separates the superelectron (below) to the vortex 
(above) dominated regime.}
\label{fig2}
\end{figure}
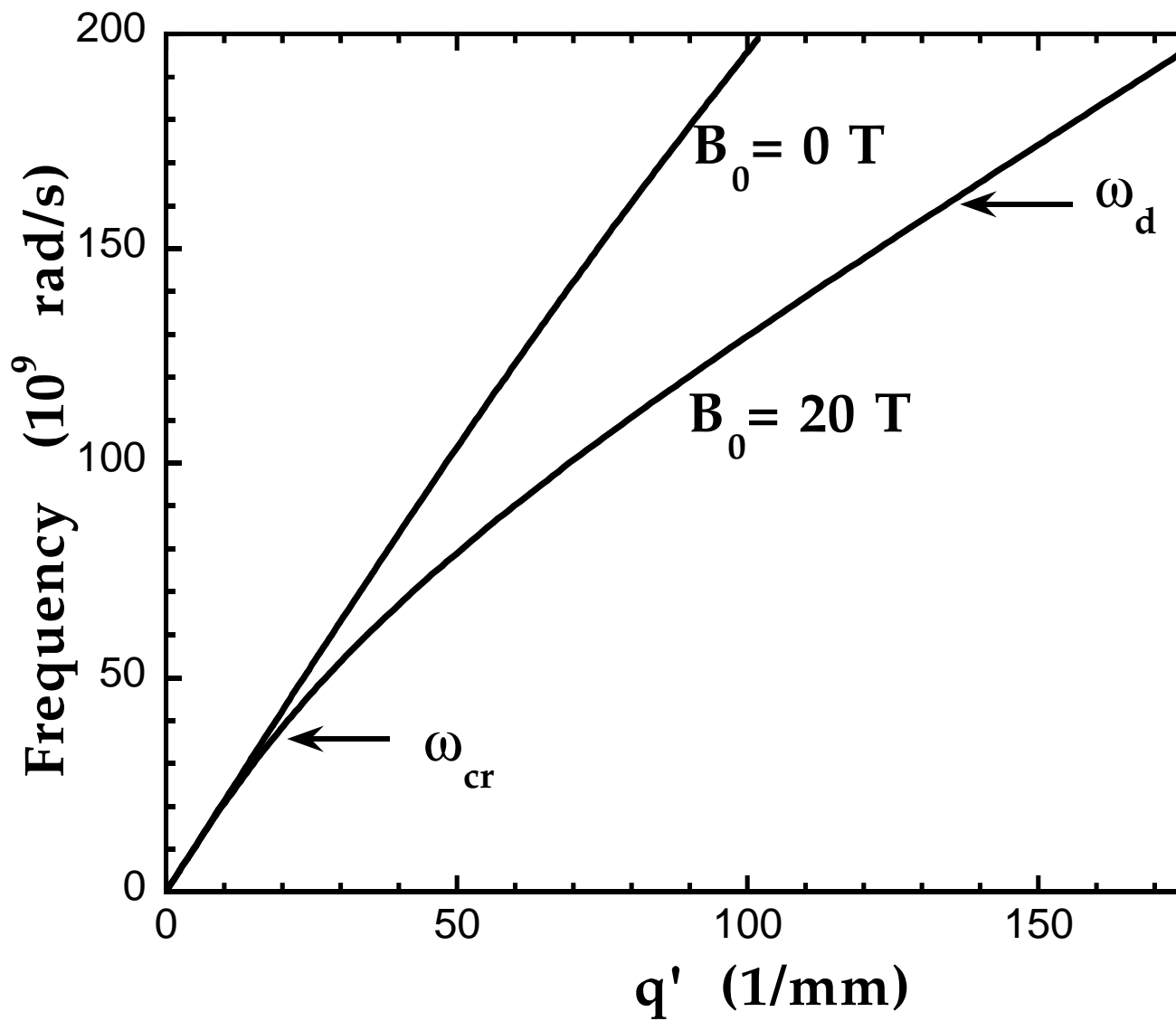
\begin{figure}
\caption{
Dispersion relation $\omega$ versus $q'$ for a $10\;nm$ $YBCO$ film.
In this frequency range and for zero magnetic field
the dispersion relation is purely optical.
For $B_0= 20 \; T$ , the modes are associated to the vortex dynamic and are
underdamped until the frequency $\omega_d$ is reached.
}
\label{fig3}
\end{figure}
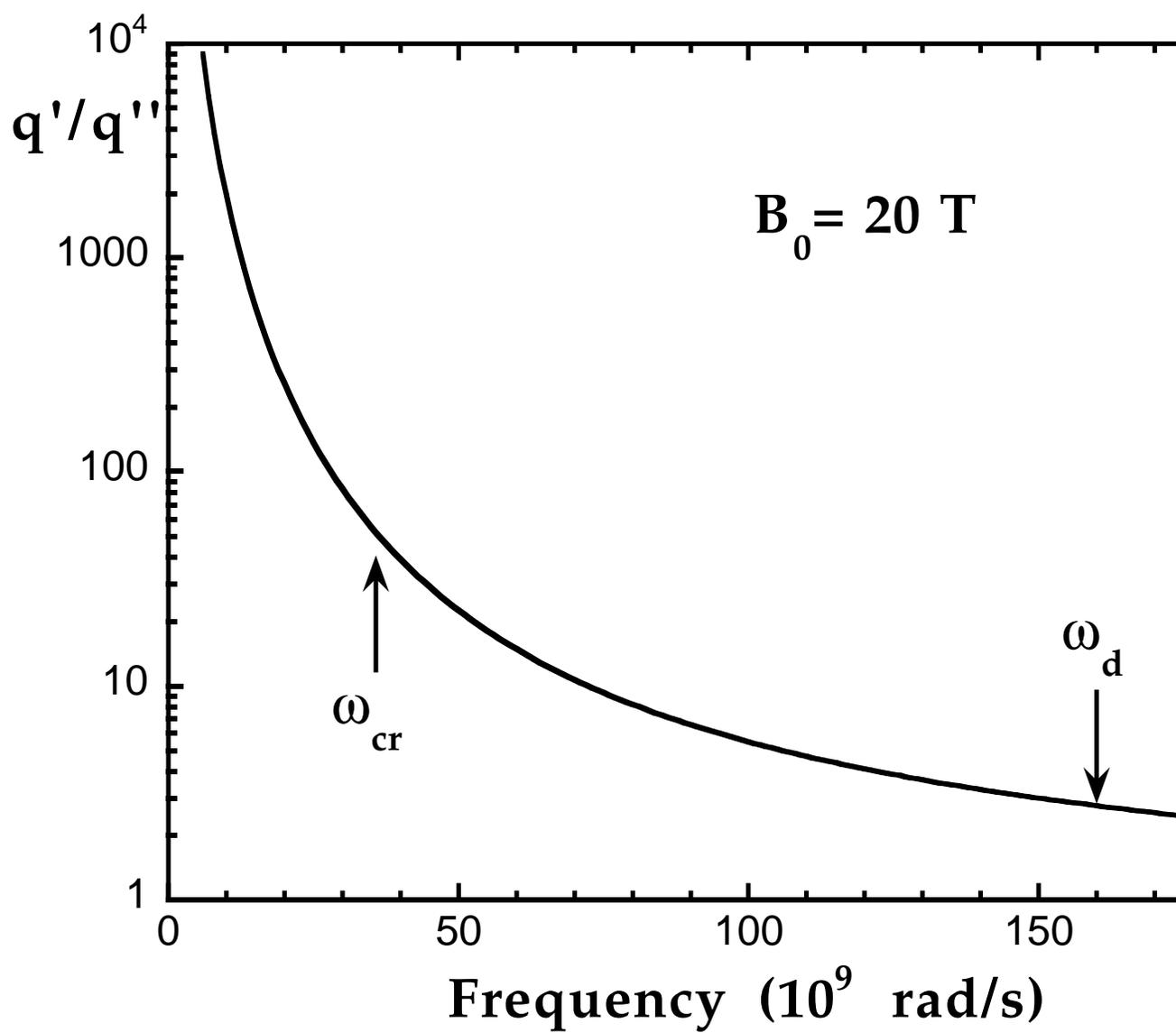
\begin{figure}
\caption{
The ratio $q'/q''$ is displayed here versus $\omega$ showing
the mode damping for the same frequency range
of Fig.3.
The ratio, although  undergoes a dramatic change in this range, is
always larger than one,
thus signaling underdamped behavior.
}
\label{fig4}
\end{figure}
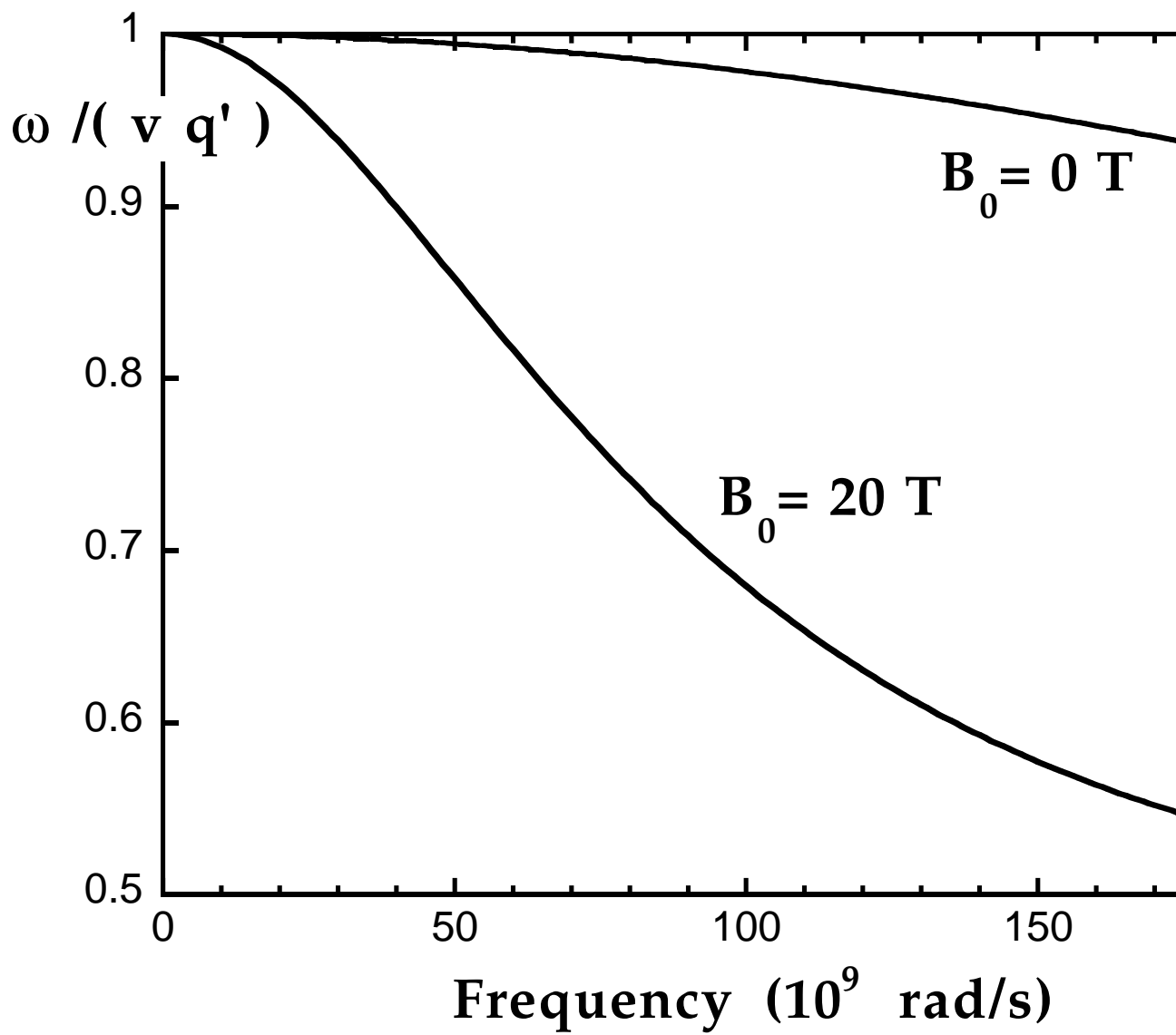
\begin{figure}
\caption{
The retardation ratio, $(\omega/q')/v$, is shown for the frequency
range of Fig.4.
For zero applied field this ratio is near one showing that
mode  is essentially  optical.
This is not case for $B_0 = 20 \; T$ whose strong deviation from one
signals coupling between the dielectric
and the superconducting film due to the presence of vortices.}
\label{fig5}
\end{figure}

\begin{references}
\bibitem{PINES} D. ~Pines and D.~Bohm, Phys. Rev. {\bf 85}, 338 (1952);
D. ~Bohm and D. ~Pines, Phys. Rev. {\bf 92}, 609 (1953).
\bibitem{BEASLEY} Tunneling and proximity effect studies
give the frequency gaps $5.0\,10^{12} \;Hz$, and $7.5\,10^{12} \;Hz$, for
$YBCO$ and $BSCCO$, respectively. See, for instance,
M.R. ~Beasley, Physica C {\bf 185}, 227 (1991).
\bibitem{MARTIN} P.C.~Martin in {\it Superconductivity} ed. R.D.
\bibitem{ANDERSON} P.W.~Anderson,  Phys. Rev. {\bf 112}, 1900 (1958);
Phys. Rev. B {\bf 130}, 439 (1963).
\bibitem{CARLSON} R.V. ~Carlson and A.M. ~Goldman, Phys. Rev. Lett.
{\bf 34}, 11 (1975).
\bibitem{ARTEMENKO} S.N. ~Artemenko and A.F. ~Volkov, Sov. Phys.
JETP {\bf 42}, 896 (1975).
\bibitem{TAMASAKU}  K.~Tamasaku, Y. ~Nakamura, and U. ~Uchida,
Phys. Rev. Lett. {\bf 69}, 1455 (1992);
 H.A.~Fertig and S.~Das~Sarma,  Phys. Rev. Lett. {\bf
65}, 1482 (1990);  Phys. Rev. B {\bf 44}, 4480 (1991);
L.N.~Bulaevskii {\it et al},  Phys. Rev. B {\bf 50} (1994) 12831;
A.M. ~Gerrits,{\it et al}, Physica C {\bf 235-240}, 1117 (1994);
\S.N.~Artemenko and A.G.~Kobelkov,  JETP Lett. {\bf
58}, 445 (1993); Physica C {\bf253} (1995) 373;
\bibitem{RITCHIE} R.H. ~Ritchie, Phys. Rev.  {\bf 106}, 874
(1957).
\bibitem{STERN} E.A. ~Stern and R.A. ~Ferrel, Phys. Rev.  {\bf 120}, 130
(1960).
\bibitem{MOOIJ} J.E.~Mooij and G.~Sch\"on, Phys. Rev. Lett. {\bf 55}, 114
(1985).
\bibitem{MIRHASHEN} B.~Mirhashem and R.~Ferrell, {\it Physica C}
{\bf 161}, 354 (1989).
\bibitem{MISHONOV} T.~Mishonov and A.~Groshev, Phys. Rev. Lett. {\bf
64}, 2199 (1990).
\bibitem{BUISSON} O.~Buisson, P.~Xavier, and J.~Richard,  Phys. Rev.
Lett. {\bf 73}, 3153 (1994); {\it Phys. Rev. Lett.} {\bf 74E}, 1493 (1995).
\bibitem{DUNMORE} F.J. ~Dunmore, D.Z. ~Liu, H.D. ~Drew, and S. ~Das Sarma,
Phys. Rev. B {\bf 52}, R731 (1995).
\bibitem{BOERSCH} H.~Boersch, J. ~Geiger, A. ~Imbush, and
N. ~Niedrig, Phys. Lett. {\bf 22}, 146 (1966).
\bibitem{FUKUI} M. ~Fukui, V.C.Y. ~So, and R. ~Normandin,
Phys. Stat. Sol. {\bf 91}, K61 (1979).
\bibitem{SARID} D.~Sarid, Phys. Rev. Lett. {\bf 47}, 1927 (1981).
\bibitem{UL} S. ~Ushioda and R. Loudon, in {\it Surface Polaritons},
edited by V.M. ~Agranovich and A.A. ~Maradudin (North-Holland P.Co.,
New York, 1982), p.573.
\bibitem{DPB} M.M. ~Doria, F.~Parage, and O. ~Buisson, Europhys. Lett.
{\bf 35}, 445 (1996).
\bibitem{GITTLEMAN} J.I.~Gittleman and B. ~Rosenblum,
Phys. Rev. {\bf 39}, 2617 (1968).
\bibitem{CAMPBELL} A.M.~Campbell ,
J. Phys. C {\bf 2}, 1492 (1969); {\bf 4}, 3186 (1971).
\bibitem{COFFEY} M.W. ~Coffey and J.R. ~Clem ,
Phys. Rev. Lett. {\bf 67}, 386 (1991).
\bibitem{BRANDT} E.H. ~Brandt, Phys. Rev. Lett. {\bf 67}, 2219 (1991).
\bibitem{FIORY} A.T. ~Fiory and A.F. ~Hebard, Phys. Rev. B {\bf 25}, 2073
(1982).
\bibitem{KOSHELEV} A.E. ~Koshelev and V.M. ~Vinokur, Physica C {\bf 173}, 465
(1991).
\bibitem{BLATTER} G. ~Blatter et al., Rev. Mod. Phys. {\bf 66},
1125 (1994).
\bibitem{BRANDT2} E.H. ~Brandt, Rep. Prog. Phys. {\bf 58}, 1465 (1995).
\bibitem{GOLOSOVSKY} M.~Golosovsky, M. ~Tsindlekht, and
D. ~Davidov, Supercond. Sci Technol. {\bf 9}, 1 (1996).
\bibitem{ANLAGE} D.H. ~Wu, J.C. ~Booth and
S.M. ~Anlage, Phys. Rev. Lett. {\bf 75}, 525 (1995).
\bibitem{HANAGURI} T. ~Hanaguri et al.,
 Physica C {\bf 235-240}, 1991 (1994).
\bibitem{KOBER} J. ~Kober et al.,
 Phys. Rev. Lett. {\bf 66}, 2507 (1990).
\end{references}
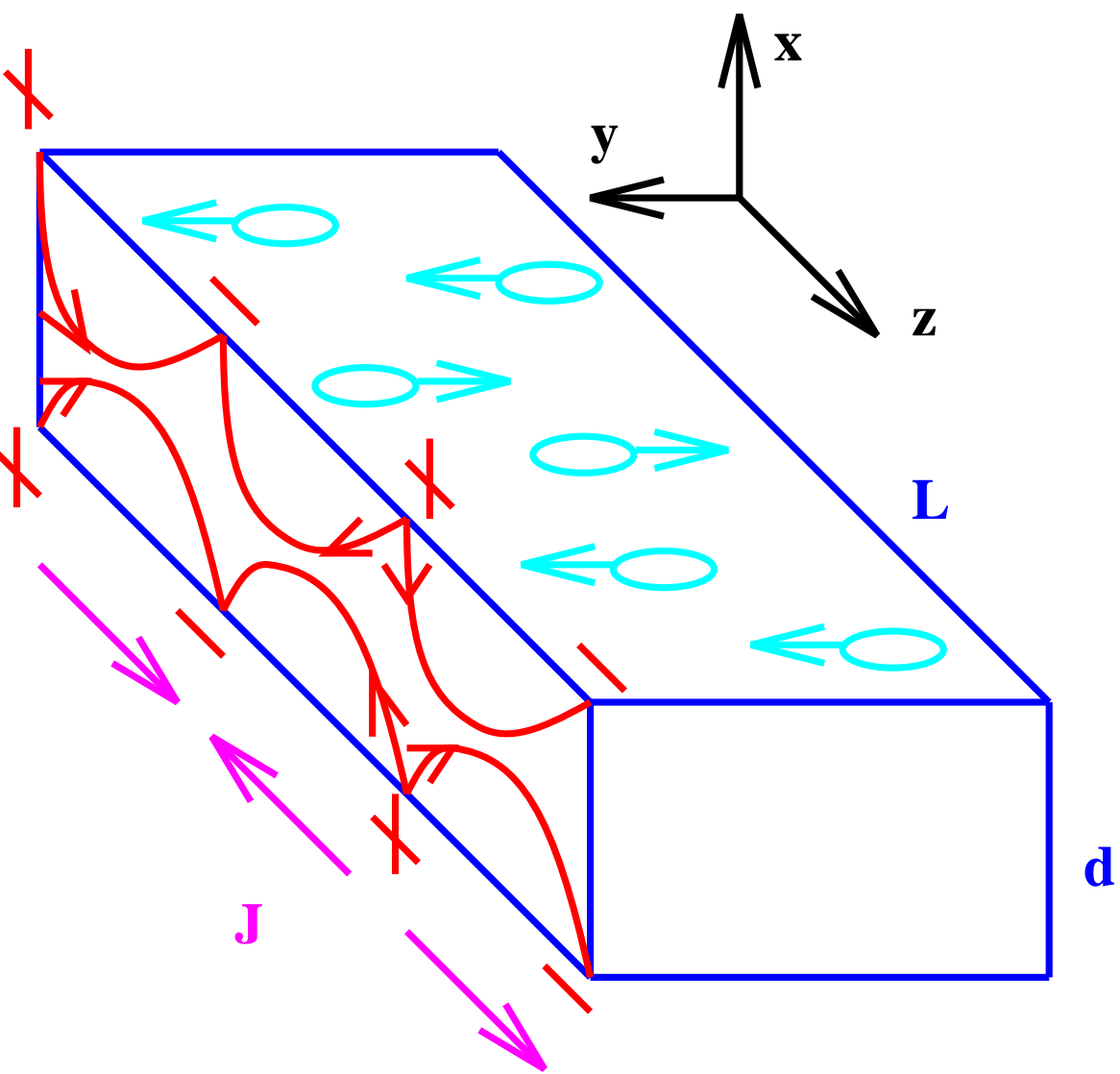
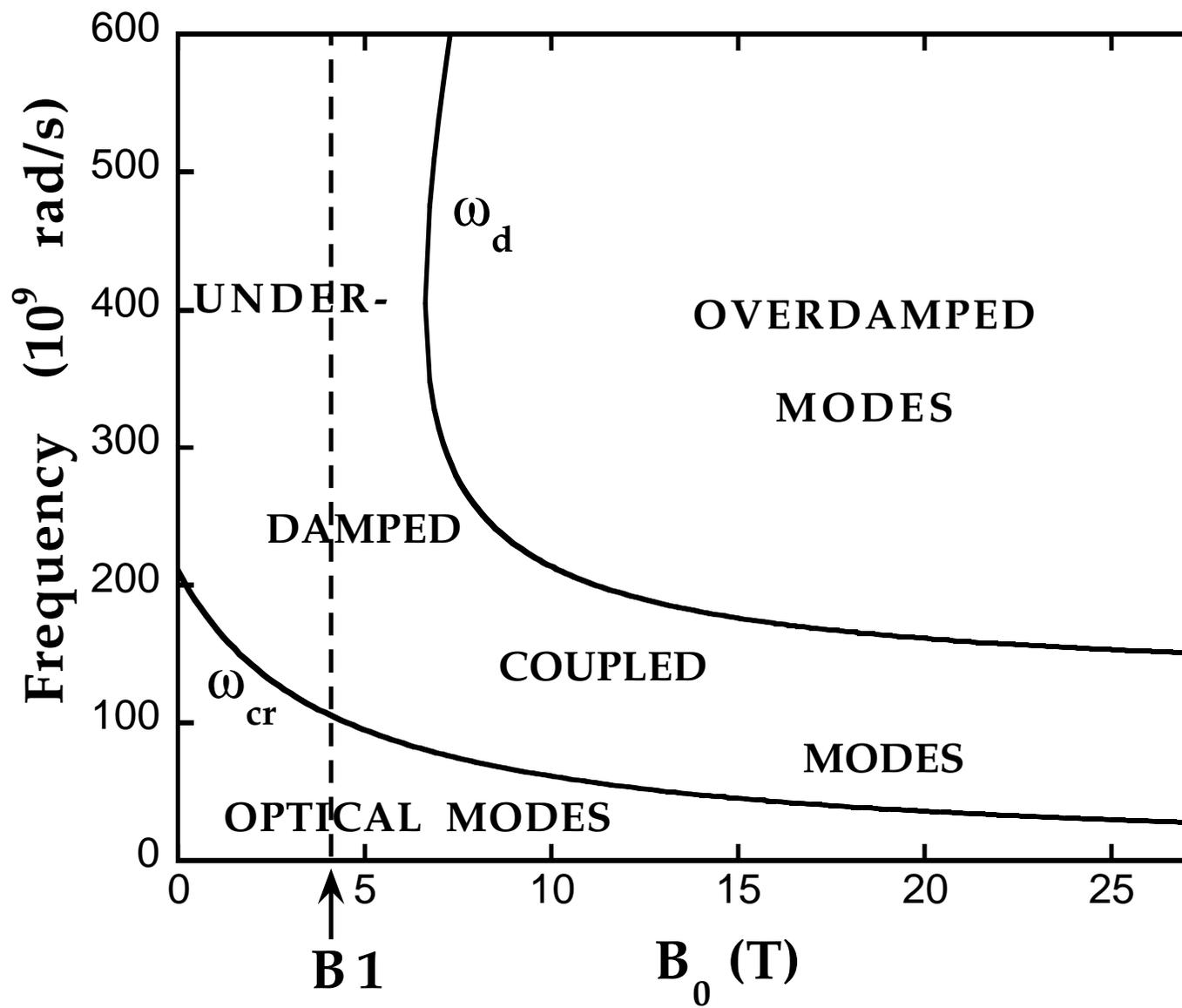
\end{document}